\documentclass[
reprint,
superscriptaddress,
preprintnumbers,
 amsmath,
 amssymb,
 prc,
]{revtex4-1}
\usepackage[english]{babel}
\usepackage[bookmarks,
             bookmarksopen = true,
             bookmarksnumbered = true,
             linktocpage,
             colorlinks = true,
             linkcolor = blue,
             urlcolor  = blue,
             citecolor = blue,
             anchorcolor = green,
             hyperindex = true,
             hyperfigures]
        {hyperref}

\usepackage{graphicx}
\usepackage{dcolumn}
\usepackage{bm}
\usepackage{float}
\usepackage{amsmath}
\usepackage{color}
\usepackage{booktabs}
\usepackage{multirow}
\definecolor{dgreen}{cmyk}{1.,0.,1.,0.2}
\definecolor{orange}{cmyk}{0.,0.353,1.,0.}

\begin{document}

\title{Neural network maximum entropy framework for distribution reconstruction in heavy-ion collisions}

\author{Qian-Ru Lin}
\affiliation{Key Laboratory of Quark and Lepton Physics (MOE) \& Institute of Particle Physics, Central China Normal University, Wuhan 430079, China}
\affiliation{Artificial Intelligence and Computational Physics Research Center, Central China Normal University, Wuhan 430079, China}

\author{Fu-Peng Li}
\email[]{fpli@fudan.edu.cn}
\affiliation{Key Laboratory of Nuclear Physics and Ion-beam Application (MOE) \& Institute of Modern Physics, Fudan University, Shanghai 200433, China}
\affiliation{Shanghai Research Center for Theoretical Nuclear Physics, NSFC and Fudan University, Shanghai 200438, China}
\affiliation{Key Laboratory of Quark and Lepton Physics (MOE) \& Institute of Particle Physics, Central China Normal University, Wuhan 430079, China}
\affiliation{Artificial Intelligence and Computational Physics Research Center, Central China Normal University, Wuhan 430079, China}

\author{Yige Huang}
\email[]{huangyige@impcas.ac.cn}
\affiliation{State Key Laboratory of Heavy Ion Science and Technology, Institute of Modern Physics, Chinese Academy of Sciences, Lanzhou 730000, China}

\author{Long-Gang Pang}
\email[]{lgpang@ccnu.edu.cn}
\affiliation{Key Laboratory of Quark and Lepton Physics (MOE) \& Institute of Particle Physics, Central China Normal University, Wuhan 430079, China}
\affiliation{Artificial Intelligence and Computational Physics Research Center, Central China Normal University, Wuhan 430079, China}

\date{\today}%

\begin{abstract}
We develop a neural-network maximum-entropy (NN+MaxEnt) framework for reconstructing probability distributions from limited observables in heavy-ion collisions. The method combines flexible neural-network representations with Shannon-entropy regularization, preserving positivity and normalization without assuming a fixed analytic form. After validation with Gaussian, Poisson, and mixed-Poisson closure tests, we apply the framework to two physics-motivated inverse problems: an effective multiplicity reconstruction constrained by functional renormalization group cumulants, used as a closure test, and the conditional jet-energy-loss distribution extracted from single-inclusive jet $R_{AA}$ data in Pb+Pb collisions at $\sqrt{s_{NN}}=2.76$~TeV. For the fRG closure test, NN+MaxEnt accurately reproduces the imposed cumulants and yields distributions consistent with conventional MaxEnt solutions. For jets, the reconstructed energy-loss distributions reproduce the measured $R_{AA}$; at an initial jet momentum $x=50~\mathrm{GeV}$, the conditional mean energy loss is $\langle\Delta p_T\rangle\simeq11.8~\mathrm{GeV}$, with a central $16\text{--}84\%$ interval of $9.0\text{--}15.0~\mathrm{GeV}$. The extracted energy-loss profile is qualitatively consistent with Bayesian MCMC and LBT results. NN+MaxEnt thus provides a flexible, less ansatz-dependent framework for regularized distribution reconstruction from observables connected to the underlying distribution through differentiable forward maps.
\end{abstract}
\maketitle

\section{Introduction}
\label{sec:intro}
Knowledge of the probability distribution of a random variable \(X\)
provides, in principle, a complete statistical description of \(X\) and
its associated uncertainty~\cite{Nagaraja2006,alma9935027391106761}.
In practice, however, the underlying distribution is rarely known and
must be inferred indirectly from finite and noisy measurements. This
challenge is especially pronounced in heavy-ion collision (HIC)
physics, where the extraction of underlying probability distributions
from experimentally accessible observables is a recurring
task~\cite{Adams2005,Gyulassy2005,Jacak2012,Shuryak2014,Heinz2002,
Sorensen:2023zkk}. Experiments generally constrain such distributions
only through a limited set of integral or moment-type observables, such
as the mean transverse momentum \(\langle p_T\rangle\), anisotropic-flow
coefficients \(v_n\), and cumulants of net-particle multiplicity
distributions~\cite{STAR2014,Adamczyk2017,Luo2017,Luo:2017faz,
Chen:2024aom,STAR:2025owm,STAR:2021yiu,STAR:2021ozh,STAR:2022gki,
STAR:2024wgy}. Inferring a probability distribution \(P(x)\) from such
limited information is mathematically ill-posed, because infinitely many
distinct distributions may reproduce the same finite set of moments or
integrated observables. A robust regularization strategy is therefore
required to connect measured observables with the underlying
distribution in a controlled manner.

The maximum-entropy principle provides a systematic prescription for
regularizing such underdetermined inference problems by selecting the
least-committal distribution consistent with the available
information~\cite{Jaynes1957,Shore1980}. In HIC fluctuation studies,
maximum-entropy methods have been used to reconstruct multiplicity and
baryon-number distributions from moment or cumulant
constraints~\cite{Pradeep2022,Hammelmann2023,Karthein2025,Huang2023}.
In jet-quenching studies, Bayesian analyses have inferred jet
energy-loss distributions from inclusive-jet suppression data using a
Gamma-distribution parametrization~\cite{He2018,He2019}. These studies
demonstrate that physically relevant distributions can be constrained
by a limited number of observables. However, existing MaxEnt
applications to HIC fluctuations are generally formulated for a
prescribed set of moment or cumulant constraints, whereas Bayesian
energy-loss extractions may inherit sensitivity to the selected
finite-dimensional analytic family.

Motivated by the rapid development of machine-learning techniques in
high-energy physics and their successful applications to HIC
problems~\cite{Baldi2014,Kasieczka2019,Andrews2020,Pang2018,
Zhou:2023pti,He:2023zin,Pang2021,Du2021,Du2022,Yang2023,Li2023,
Liu2023,Mallick2022,Carrazza2019,Pang:2024kid,Boehnlein:2021eym},
neural networks have increasingly been used as flexible representations
for ill-posed inverse problems. Closely related to the present approach,
MEP-Net reconstructs an unknown probability density from moment or
feature constraints using a neural-network representation and an
entropy-related regularization~\cite{YangMEPNet2025}. Neural networks
have also been employed to reconstruct spectral functions through
differentiable integral forward models and automatic
differentiation~\cite{Wang:2021jou}, within the broader context of
physics-driven inverse problems in QCD~\cite{Aarts:2025gyp}.

In this work, we develop a neural-network-based,
maximum-entropy-regularized framework, referred to as NN+MaxEnt, for
reconstructing probability distributions and conditional densities from
limited observables without imposing a fixed analytic family. The
central ingredient is a numerically differentiable and computationally
tractable forward map \(F[P]\) connecting the unknown distribution to
the constrained observable. The distribution is represented by a
positive neural network and optimized using observable constraints,
normalization, and Shannon-entropy regularization, with an additional
smoothness penalty when required. Relative to moment-constrained neural
reconstruction, the present formulation accommodates general
nonlocal integral forward operators and conditional probability
densities. This extension is essential for the jet application below,
where the measured inclusive suppression constrains the energy-loss
distribution only indirectly through a convolution with the
proton--proton spectrum.

We first validate the method through Gaussian, Poisson, and mixed-Poisson
closure tests. We then consider two HIC applications with distinct
roles. The first is a physics-motivated closure reconstruction based on
rescaled functional-renormalization-group (fRG) cumulants; because the
volume, experimental acceptance, and baryon-to-proton mapping are not
fixed, this reconstruction is not interpreted as a quantitative
prediction of an experimentally measured proton multiplicity
distribution. The second is the inverse reconstruction of an effective
conditional jet-energy-loss distribution,
\[
W_{AA}(\Delta p_T \mid p_T+\Delta p_T),
\]
from single-inclusive jet \(R_{AA}\) data in Pb+Pb collisions at
\(\sqrt{s_{NN}}=2.76\)~TeV~\cite{CMS2011,ATLAS2013,ATLAS2015,ALICE2014}.
The latter problem involves a nonlocal integral forward map and is
therefore substantially more underdetermined than a finite set of moment
constraints. The reconstructed jet-energy-loss distribution is compared
qualitatively with the Bayesian MCMC extraction of Ref.~\cite{He2018}
and with LBT calculations. Our purpose is not to claim a unique
microscopic inversion of inclusive \(R_{AA}\), but to demonstrate a
flexible and less ansatz-dependent regularized reconstruction that makes
the role of the forward map and regularization explicit.

The remainder of this paper is organized as follows.
Section~\ref{sec:method} describes the NN+MaxEnt framework.
Section~\ref{sec:res} presents the benchmark closure tests and the two
physics applications to proton-number and jet-energy-loss distributions.
Section~\ref{sec:sum} summarizes the main results and discusses future
directions.

\section{Method}
\label{sec:method}
\begin{figure*}[htbp]
    \centering
    \includegraphics[width=1\textwidth]{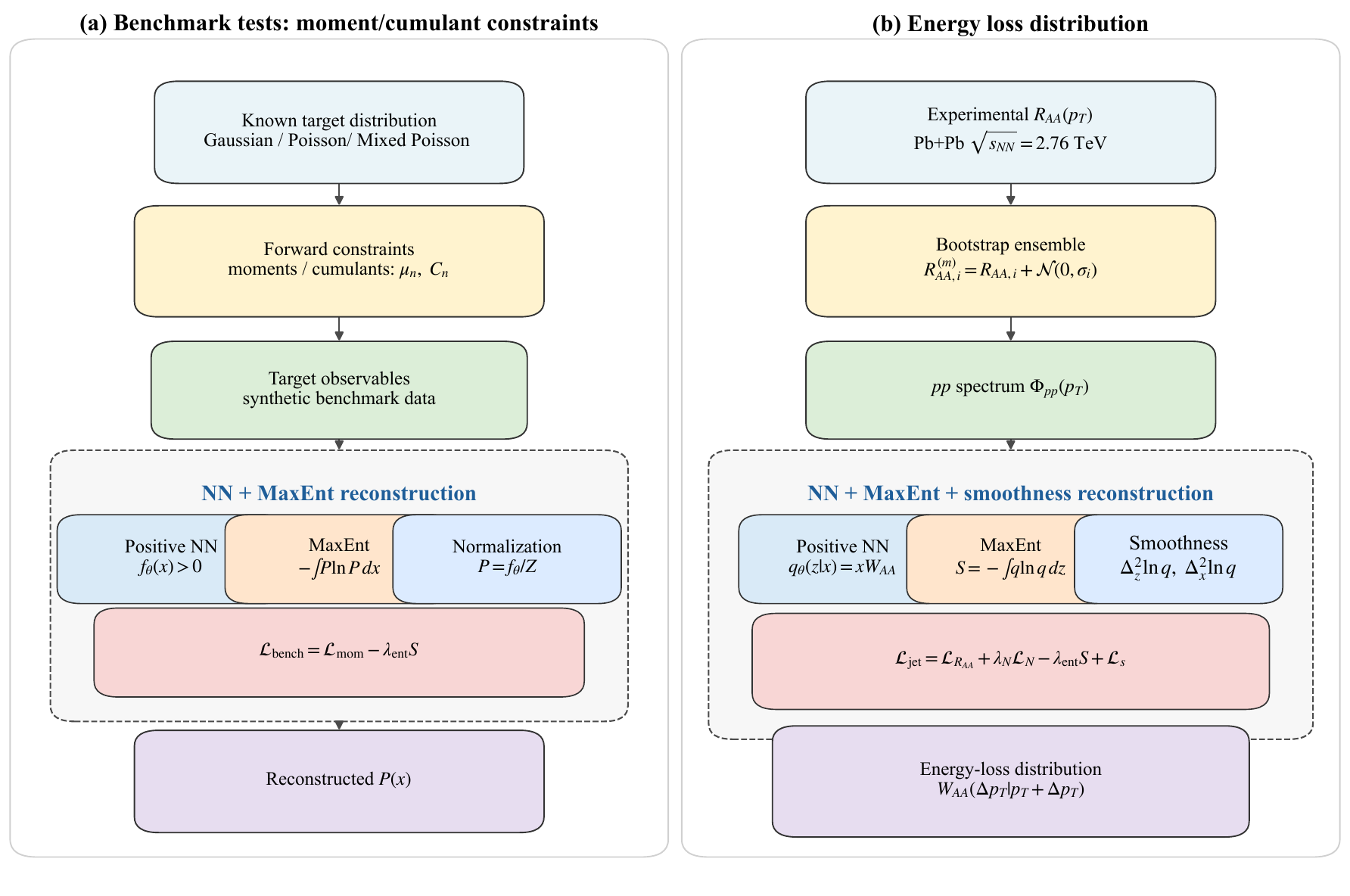}
    \caption{Workflow diagram of the NN+MaxEnt framework. \textbf{(a)} Benchmark tests: the target distribution generates exact moments or cumulants, which are used to train the NN+MaxEnt model through the moment loss $\mathcal{L}_{\text{mom}}$. \textbf{(b)} Energy-loss reconstruction from experimental $R_{AA}(p_{\mathrm{T}})$ data. }
    \label{fig:workflow}
\end{figure*}

The NN+MaxEnt workflow used in this study is summarized in Fig~\ref{fig:workflow}. We represent the probability density with a deep neural network (NN) containing four residual (ResNet) blocks~\cite{He2015}, each with 64 hidden units. Each residual block contains two fully connected layers with $\tanh$ activation and a skip connection. A positive output activation enforces positivity during optimization. We train the network with the Adam optimizer~\cite{Kingma2014}, using an initial learning rate of $10^{-3}$ that decays by a factor of $0.95$ every 500 epochs. Training is run for up to 20,000 epochs with an early-stopping patience of 2,000 epochs. All models are implemented in PyTorch~\cite{PyTorch}. 

\subsection{Maximum Entropy Principle}
\label{subsec:maxent}

The maximum-entropy (MaxEnt) principle, introduced by Jaynes based on the
Shannon entropy~\cite{Shannon1948,Jaynes1957}, provides a systematic
prescription for inferring a probability distribution from incomplete
information. For a normalized distribution \(P(x)\), the Shannon entropy
is
\begin{equation}
    S[P]=-\int dx\,P(x)\ln P(x).
\end{equation}
When the available information can be expressed as expectation-value
constraints,
\begin{equation}
    \int dx\,P(x)O_i(x)=\langle O_i\rangle ,
\end{equation}
the MaxEnt distribution is obtained by maximizing \(S[P]\) while
satisfying these constraints and the normalization condition
\(\int dx\,P(x)=1\).

In the conventional Lagrange-multiplier formulation, the constrained
optimization is written as
\begin{equation}
\begin{split}
    \mathcal{J}[P] ={}& S[P]
    -\lambda_0\left(\int dx\,P(x)-1\right) \\
    &-\sum_i\lambda_i
    \left[
        \int dx\,P(x)O_i(x)-\langle O_i\rangle
    \right],
\end{split}
\end{equation}
where \(\lambda_i\) are Lagrange multipliers. The stationarity condition
\(\delta\mathcal{J}/\delta P(x)=0\) leads to
\begin{equation}
    P_{\rm MaxEnt}(x)
    =
    \frac{1}{Z}
    \exp\left[-\sum_i\lambda_i O_i(x)\right],
    \label{eq:lag_maxent}
\end{equation}
where \(Z\) ensures normalization. The multipliers \(\lambda_i\) are
determined by requiring the reconstructed distribution to reproduce the
imposed constraints. For moment constraints, for example,
\(O_i(x)=x^i\), and Eq.~(\ref{eq:lag_maxent}) gives the conventional
MaxEnt distribution used as a reference in the proton-number
reconstruction.

The NN+MaxEnt approach adopted in this work differs from the conventional
construction in how the distribution and constraints are implemented.
Instead of imposing the exponential form of Eq.~(\ref{eq:lag_maxent}),
we represent the unknown distribution by a positive neural network and
incorporate the observable constraints through differentiable penalty
terms in the training loss. The Shannon entropy then acts as a
regularizer that favors higher-entropy solutions among those compatible
with the constrained observables. Because the constraints are enforced
numerically rather than as exact equality constraints, NN+MaxEnt is not
mathematically identical to the classical Lagrange-multiplier solution.
Throughout this work, the term NN+MaxEnt therefore denotes a neural,
entropy-regularized implementation of the maximum-entropy principle.

\subsection{Benchmark tests}
Figure~\ref{fig:workflow}(a) shows the benchmark-test workflow, in which the target observables are moments and cumulants. For a given distribution $P(x)$, its raw moments $\mu'$ and central moments $\mu$ are defined as
\begin{align}
    \mu'_{n}& = \int d x\, x^{n} P(x),
    \\ \nonumber
    \mu_{n} &= \int d x\, (x - \mu'_{1})^{n}\,P(x),
    \qquad n\geq 2,
\end{align}
For discrete distributions, the integrals are replaced by sums over the
integer support. The cumulants $C_{n}$ can be expressed by these moments:
\begin{align}
    C_{1} &= \mu'_{1},  \quad C_{2} = \mu_{2}, \\ \nonumber
    C_{3} &= \mu_{3},   \quad  C_{4} = \mu_{4} - 3\mu_{2}^{2}.
\end{align}
In the benchmark tests, the input vector is chosen as
\begin{equation}
    \mathbf{M} \equiv [\mu'_{1}, \mu_{2}, \mu_{3}, \mu_{4}, C_{4}] .
\end{equation}
The central moments $\mu_2$, $\mu_3$, and $\mu_4$ characterize the width, asymmetry, and fourth-order structure of the distribution, while the fourth cumulant $C_4$ isolates the connected fourth-order contribution by subtracting the disconnected Gaussian term \(3\mu_2^2\). Once $\mu_2$ is fixed, however, $\mu_4$ and $C_4$ are not independent constraints. We retain both in the numerical loss so that the optimization places explicit emphasis on the total fourth-order moment and on its connected component. Their simultaneous inclusion should therefore be understood as a numerical weighting choice rather than as additional independent information. For target observables with large magnitudes ($\mu'_{1} \gtrsim 20$), we use relative errors:
\begin{equation}
    \mathcal{L}_{\text{mom}} = \left[\sum_{n=1}^{5} \left(\frac{M_{n}^{\text{NN}}}{M_{n}^{\text{input}}} - 1\right)^{2}\right]^{1/2},
\end{equation}
where $M_n$ denotes the $n$-th component of $\mathbf{M}$. For small-magnitude targets, absolute errors are used instead. Notably, the benchmark loss does \emph{not} include a smoothness penalty; the entropy term alone is sufficient to regularize these low-dimensional test cases. The total loss is simply
\begin{equation}
    \mathcal{L}_{\text{total}} =  \mathcal{L}_{\text{mom}} - \lambda_{ent} S, \quad \lambda_{ent} = 20.
\end{equation}

\subsection{Proton-number reconstruction from fRG cumulants}
\label{subsec:frg_proton}

As a physics-motivated test of NN+MaxEnt, we use fluctuation observables
obtained from fRG calculations. The
fRG provides a nonperturbative description of QCD thermodynamics and
phase structure at finite temperature and baryon chemical potential
~\cite{Fu:2021oaw,Fu:2019hdw}. Higher-order baryon-number fluctuations are
particularly sensitive to changes in the QCD phase structure and may
develop enhanced or nonmonotonic behavior near a critical region
~\cite{Stephanov2009,Luo2017}. Such structures, however, are not unique
signatures of a critical endpoint and may also arise from a rapid
crossover.

The baryon-number susceptibilities are defined by derivatives of the
dimensionless pressure,
\begin{equation}
    \chi_n^B(T,\mu_B)
    =
    \frac{\partial^n [p(T,\mu_B)/T^4]}
    {\partial(\mu_B/T)^n}.
    \label{eq:chi_frg}
\end{equation}
In the grand-canonical ensemble, they are related to the cumulants of the
net-baryon-number distribution through
\begin{equation}
    C_n^B = VT^3\,\chi_n^B ,
    \label{eq:cn_chin}
\end{equation}
where \(V\) is the system volume. 

In the present study, the first four fRG cumulants at selected collision
energies are used as input constraints for a physics-motivated closure
test. Since the overall factor \(VT^3\), experimental acceptance, and the
mapping from baryon-number to proton-number fluctuations are not fixed,
the input cumulants are rescaled for numerical conditioning. The same
preprocessed cumulant set is used for both NN+MaxEnt and the conventional
Lagrange-multiplier MaxEnt method. We retain the notation \(P(N_p)\) for
the reconstructed multiplicity variable used in the numerical analysis,
but its absolute scale is not fixed by the fRG susceptibilities alone.
The reconstruction should therefore be interpreted as an effective
cumulant-constrained multiplicity distribution and a closure benchmark,
rather than as a quantitative prediction of the experimentally measured
proton-number distribution.

\subsection{Jet Energy loss distribution}
\label{subsec:loss}
Figure~\ref{fig:workflow}(b) shows the workflow for energy-loss distribution reconstruction. We use the NN to reconstruct the energy-loss distribution \(W_{AA}(\Delta p_T|p_T+\Delta p_T)\). To facilitate learning and impose normalization, the network uses the dimensionless variable $z=\Delta p_{\mathrm{T}}/x$ together with the initial jet momentum $x$, where $x = p_T+\Delta p_T$. The NN output is
\begin{equation}
    q_\theta(z|x)= \mathrm{NN}_\theta(x,z) .
    \label{eq:qtheta}
\end{equation}
The relation between $q_\theta$ and $W_{AA}$ follows from a change of variables in the first argument of the conditional density. Holding the conditioning variable $x$ fixed gives $d\Delta p_{\mathrm{T}}=x\,dz$. Probability conservation then requires
\begin{equation}
    W_{AA}(\Delta p_{\mathrm{T}}|x)\,d\Delta p_{\mathrm{T}}
    = q_\theta(z|x)\,dz ,
    \label{eq:prob_conservation}
\end{equation}
and therefore
\begin{equation}
\begin{split}
    q_\theta(z|x)
    &=x\,W_{AA}(\Delta p_{\mathrm{T}}=xz|x),\\
    W_{AA}(\Delta p_{\mathrm{T}}|x)
    &=\frac{1}{x}\,
    q_\theta\!\left(\frac{\Delta p_{\mathrm{T}}}{x}\,\vert\,x\right).
\end{split}
    \label{eq:q_waa_relation}
\end{equation}
The nuclear modification factor $R_{AA}$ is then computed as
\begin{equation}
    R_{AA}(p_T)=
    \frac{1}{\Phi_{pp}({p_T})}\int_0^\infty d\Delta p_{\mathrm{T}}\,
    \Phi_{pp}(x)
    W_{AA}(\Delta p_{\mathrm{T}}|x),
    \label{eq:aa_spectrum_delta}
\end{equation}
Equivalently, using $x=p_T/(1-z)$ and $d\Delta p_{\mathrm{T}}=p_T\,dz/(1-z)^2$, together with Eq.~(\ref{eq:q_waa_relation}), the predicted nuclear modification factor is then
\begin{equation}
\begin{split}
    R_{AA}^{\rm pred}(p_T)
    =\frac{1}{\Phi_{pp}(p_T)}
    \int_0^1 dz\,
    &\frac{\Phi_{pp}\!\left(p_T/(1-z)\right)}{1-z}\\
    &\times q_\theta\left(z\,\,\vert\,\,p_T/(1-z)\right).
\end{split}
    \label{eq:raa_integral}
\end{equation}
In practice, the upper integration region corresponding to \(z\to1\) is regulated by imposing a finite maximum value of \(x\), or equivalently a cutoff \(z_{\max}<1\). The omitted contribution is numerically negligible due to the steeply falling \(pp\) spectrum. The remaining integral is evaluated by Gauss--Legendre quadrature.

Equation~(\ref{eq:raa_integral}) also makes explicit the ill-posed nature
of the inverse problem: the inclusive one-dimensional observable
\(R_{AA}(p_T)\) constrains an underlying conditional density that depends
on both the fractional loss \(z\) and the initial momentum \(x\).
Consequently, the data do not determine a unique functional form of
\(q_\theta(z|x)\). The entropy and smoothness terms below should therefore
be viewed as regularization criteria that select a representative smooth
solution compatible with the measured suppression, rather than as
additional experimental information.

The jet loss used in training is
\begin{equation}
\mathcal{L}_{\rm tot}
=
\mathcal{L}_{R_{AA}}
+\lambda_N \mathcal{L}_N
-\lambda_{\rm ent} S
+\mathcal{L}_s ,
\label{eq:jet_loss}
\end{equation}
with
\begin{align}
\mathcal{L}_{R_{AA}}
&=
\frac{1}{N_{\rm data}}
\sum_{i=1}^{N_{\rm data}}
\left[
R_{AA,i}^{(m)}-R_{AA,i}^{\rm pred}
\right]^2 ,
\\
\mathcal{L}_N
&=
\left\langle
\left(
\int_0^1 q_\theta(z|x)\,dz-1
\right)^2
\right\rangle_x ,
\\
S
&=
-\left\langle
\int_0^1 q_\theta(z|x)\ln q_\theta(z|x)\,dz
\right\rangle_x ,
\\
\mathcal{L}_s
&=
\lambda_z
\left\langle
\left(\Delta_z^2\ln q_\theta\right)^2
\right\rangle
+
\lambda_x
\left\langle
\left(\Delta_x^2\ln q_\theta\right)^2
\right\rangle .
\label{eq:norm_entropy_loss}
\end{align}

We use an unweighted mean-squared-error loss for \(R_{AA}\). Thus, the
measured momentum points enter the central reconstruction with equal
weight, while their quoted pointwise uncertainties are propagated through
the bootstrap replicas. The bootstrap procedure is used for uncertainty
propagation and is not equivalent to a covariance-weighted \(\chi^2\)
fit. The term \(\mathcal{L}_N\) enforces normalization, while
\(\mathcal{L}_s\) suppresses nonphysical oscillations by penalizing second
finite differences of \(\ln q_\theta\) along \(z\) and \(x\)
~\cite{Tikhonov1963,Tikhonov1977}. For fixed \(x\), maximizing the Shannon entropy of a normalized
\(q(z|x)\) on \(z\in[0,1]\) without the \(R_{AA}\) constraint gives the
uniform solution \(q(z|x)=1\), corresponding through
Eq.~(\ref{eq:q_waa_relation}) to \(W_{AA}(\Delta p_T|x)=1/x\). Thus, the entropy term alone does not favor
vanishing energy loss, and the uniform solution is also compatible with
the smoothness penalty. We use
\(\lambda_N=20\),
\(\lambda_{\rm ent}=10^{-3}\),
\(\lambda_z=5\times10^{-4}\), and
\(\lambda_x=10^{-4}\).
A one-at-a-time sensitivity scan of these parameters is presented in
Appendix~\ref{sec:lambda}.

\subsection{Ensemble Uncertainty Quantification}
\label{subsec:ensemble}

To propagate the experimental input uncertainty and reconstruction
variability, we train an ensemble of \(N_{\rm model}=50\) neural
networks. Each member is trained on a bootstrap replica,
\(R_{AA,i}^{(m)}=R_{AA,i}+\mathcal{N}(0,\sigma_i)\), with fixed bin
centers and an independent random initialization. The resulting
\(16\)--\(84\%\) ensemble band therefore reflects the propagated
pointwise experimental uncertainties and the variation associated with
random network initialization. It does not include systematic
uncertainties associated with the regularization parameters, the \(pp\)
baseline, or other modeling choices, and should not be interpreted as the
full uncertainty of the inverse reconstruction.

\section{Results and discussion}
\label{sec:res}
\subsection{Gaussian and Poisson Distributions}
\begin{figure}[htbp]
    \centering
    \includegraphics[width=\columnwidth]{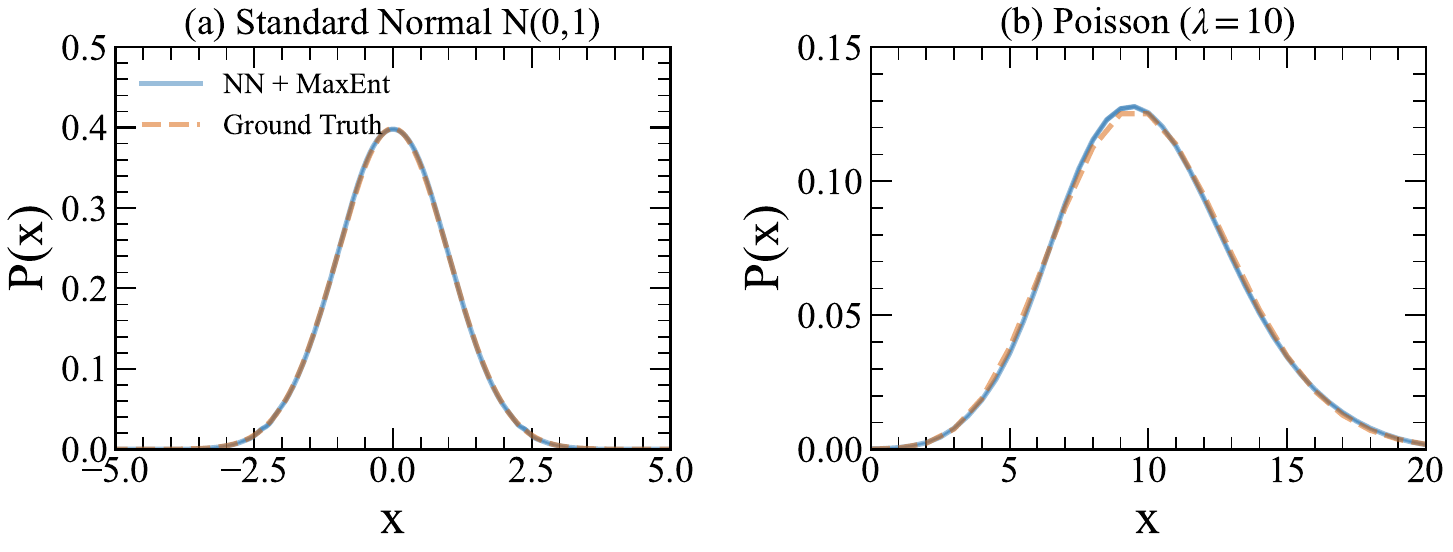}
    \caption{Closure tests for (a) a Gaussian distribution and (b) a Poisson distribution. The blue band denotes the NN+MaxEnt ensemble result, and the orange dashed line denotes the true distribution.}
    \label{fig:gaussian_pmf}
\end{figure}

We first test a Gaussian distribution, $P(x) = (2\pi\sigma^{2})^{-1/2}\exp[-(x-\mu)^{2}/(2\sigma^{2})]$, and a Poisson distribution, $P(x;\lambda)=\lambda^{x}e^{-\lambda}/x!$. Figure~\ref{fig:gaussian_pmf} shows that the reconstructed distributions agree with the true Gaussian and Poisson targets within the ensemble uncertainty bands. These closure tests show that, for target distributions compatible with
the imposed moment or cumulant constraints, NN+MaxEnt can reproduce the
dominant features of the distribution, including the peak location and
the overall tail behavior, without imposing a fixed analytic ansatz.

\subsection{Poisson and Mixed-Poisson Distributions}
\begin{figure}[htbp]
    \centering
    \includegraphics[width=0.48\textwidth]{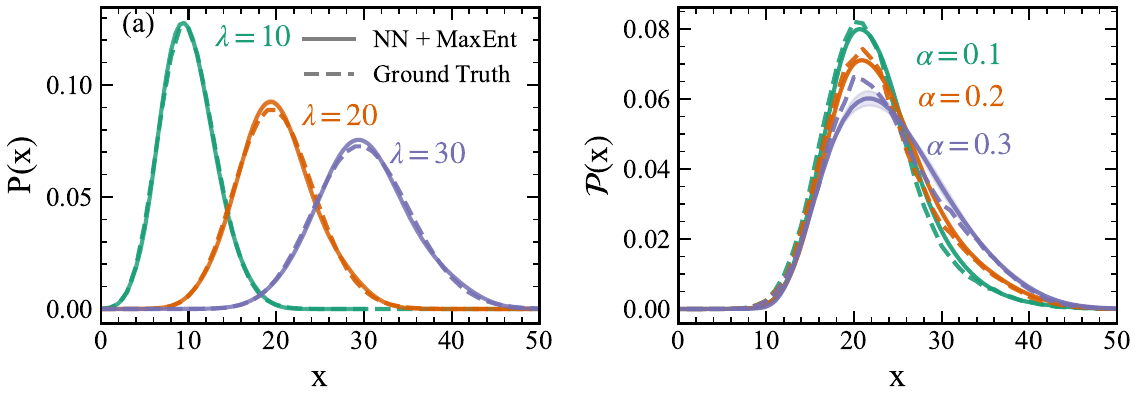}
    \caption{
        Benchmark reconstruction of Poisson-type distributions.
        (a) Poisson probability mass functions with \(\lambda=10,20,\) and \(30\).
        (b) Mixed-Poisson distributions with weight parameters \(\alpha=0.1,0.2,\) and \(0.3\).
        In both panels, the NN+MaxEnt reconstructions agree well with the corresponding target distributions, demonstrating that the framework can reproduce asymmetric and non-Gaussian probability distributions from moment or cumulant constraints.
    }
    \label{fig:poisson_mixed}
\end{figure}

We further test NN+MaxEnt with Poisson-type distributions under different conditions. Figure~\ref{fig:poisson_mixed}(a) shows the Poisson distributions with different mean values, \(\lambda=10,20,\) and \(30\). The NN+MaxEnt reconstructions reproduce the peak positions and widths of the target distributions, indicating that the method remains stable as the characteristic scale of the distribution changes. As the most stringent test, we consider a two-component mixture of Poisson distributions:
\begin{equation}
\mathcal{P}(x) = \frac{1}{1+\alpha} \, \mathrm{P}(x;20) + \frac{\alpha}{1+\alpha} \, \mathrm{P}(x;30),
\end{equation}
where $\mathrm{P}(x;\lambda) = \lambda^x e^{-\lambda}/x!$ denotes the Poisson probability mass function, and $\alpha \ge 0$ controls the relative weight of the components. Figure~\ref{fig:poisson_mixed}(b) shows the reconstruction of this mixed distribution for weight parameters $\alpha = 0.1, 0.2,$ and $0.3$. As \(\alpha\) increases, the distribution shape becomes broader and increasingly skewed. The NN+MaxEnt results follow this trend and remain consistent with the target distributions, demonstrating that the framework can faithfully capture the richer non-Gaussian features that emerge from component mixing, beyond what a single Poisson distribution exhibits.

\subsection{fRG cumulant closure reconstruction}
We next apply NN+MaxEnt to the effective multiplicity reconstruction
\(P(N_p)\) using the rescaled fRG cumulants described in
Sec.~\ref{subsec:frg_proton}. As emphasized above, this is a
physics-motivated closure test rather than a quantitative prediction of
the experimentally measured proton-number distribution. The same rescaled cumulant inputs are used
for both the NN+MaxEnt and conventional Lagrange-multiplier MaxEnt
reconstructions, allowing a direct comparison between the two methods.

\begin{figure}
    \centering
    \includegraphics[width=0.4\textwidth]{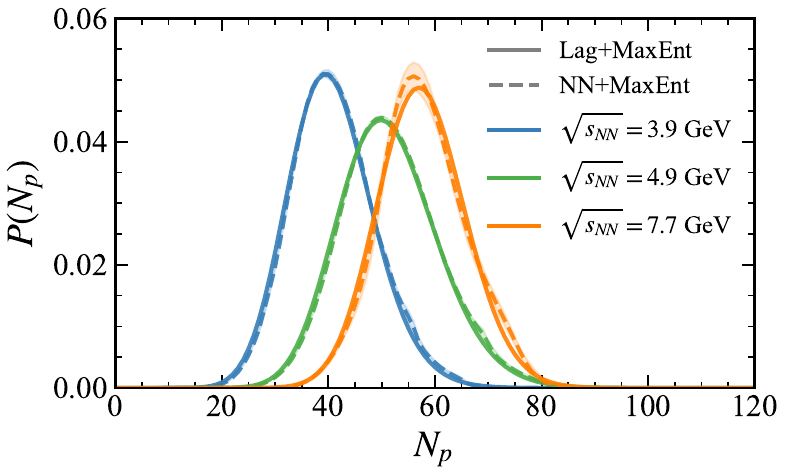}
    \caption{Effective multiplicity distributions reconstructed from the same rescaled fRG cumulants with NN+MaxEnt and the Lagrange-multiplier MaxEnt method. The comparison is used as a cumulant-closure benchmark and is not a quantitative prediction of the measured proton-number distribution.}
    \label{fig:proton_pdf}
\end{figure}

Figure~\ref{fig:proton_pdf} compares the reconstructed effective
multiplicity distributions at three collision energies. Overall, the
NN+MaxEnt and Lagrange-multiplier MaxEnt results show similar distribution
shapes, while NN+MaxEnt reproduces the imposed cumulants with high
accuracy. Within this closure setup, the comparison shows that a flexible
neural representation can recover a high-entropy,
cumulant-constrained distribution without imposing the specific
exponential form of the conventional solution.

\begin{table}[htbp]
\centering
\begin{tabular}{lccc}
\\
\toprule
Energy & Method & \(\max_{n\leq4}\delta C_n\) & \(D_{L_1}\) \\
\midrule
\hline
\multirow{2}{*}{\(\sqrt{s_{NN}}=3.9\)~GeV}
& Lag+MaxEnt & \(2.14\times10^{-1}\) & --- \\
& NN+MaxEnt  & \(3.17\times10^{-4}\) & \(4.21\times10^{-2}\) \\
\addlinespace
\hline
\multirow{2}{*}{\(\sqrt{s_{NN}}=4.9\)~GeV}
& Lag+MaxEnt & \(6.44\times10^{-12}\) & --- \\
& NN+MaxEnt  & \(1.23\times10^{-4}\) & \(3.44\times10^{-2}\) \\
\addlinespace
\hline
\multirow{2}{*}{\(\sqrt{s_{NN}}=7.7\)~GeV}
& Lag+MaxEnt & \(3.37\) & --- \\
& NN+MaxEnt  & \(8.99\times10^{-3}\) & \(6.80\times10^{-2}\) \\
\bottomrule
\end{tabular}
\caption{Quantitative comparison between the conventional Lagrange-multiplier MaxEnt method and the NN+MaxEnt reconstruction for selected collision energies. Here \(\max_{n\leq4}\delta C_n\) denotes the maximum relative error among the first four cumulants, with \(\delta C_n=|C_n^{\rm pred}-C_n^{\rm input}|/|C_n^{\rm input}|\). The quantity \(D_{L_1}\) measures the distance between the Lagrange-multiplier MaxEnt and NN+MaxEnt reconstructed distributions.}
\label{tab:validation_energies}
\end{table}
To quantify the consistency between the NN+MaxEnt and conventional Lagrange-multiplier MaxEnt reconstructions, Table~\ref{tab:validation_energies} summarizes the maximum relative cumulant error and the \(L_1\) distance for selected collision energies. The small cumulant errors of the NN+MaxEnt results indicate that the neural-network reconstruction accurately reproduces the imposed cumulant constraints. The \(D_{L_1}\) values further show that the reconstructed distributions are close to the corresponding Lagrange-multiplier MaxEnt solutions.

As shown in Table~\ref{tab:validation_energies}, the Lagrange-multiplier MaxEnt reconstruction accurately reproduces the input cumulants at \(\sqrt{s_{NN}}=4.9\)~GeV but shows sizable deviations for the selected setups at \(\sqrt{s_{NN}}=3.9\) and \(7.7\)~GeV. For the same preprocessed inputs, NN+MaxEnt reproduces all four cumulants with sub-percent accuracy at these energies. This comparison demonstrates the practical numerical flexibility of the neural parameterization in the present closure tests. The observed deviations of the particular Lagrange-multiplier reconstruction used here should not be interpreted as an intrinsic limitation of the classical MaxEnt principle or as a general proof of superiority of the neural implementation.

\subsection{Jet energy-loss distribution from $R_{AA}$ data for QGP transport}
\label{subsec:application}
Jet quenching—the suppression of high-transverse-momentum jets in heavy-ion collisions—provides one of the most direct probes of the quark–gluon plasma (QGP) created at the Relativistic Heavy Ion Collider and the LHC. The single-inclusive jet nuclear modification factor $R_{AA}(p_T)$ encodes the attenuation of jet spectra, but by itself does not uniquely determine the event-by-event jet energy-loss distribution. Inferring such a distribution is therefore a classic ill-posed inverse problem. A regularized reconstruction can nevertheless provide information beyond the mean suppression and offers a useful basis for comparison with microscopic descriptions of jet--medium interactions. 

Here we apply the NN+MaxEnt framework to single-inclusive jet $R_{AA}$ data in Pb+Pb collisions at $\sqrt{s_{NN}}=2.76$~TeV, for the $0$--$10\%$ centrality class. The jet kinematic setup and jet definition follow those used in the Bayesian extraction of Ref.~\cite{He2018}. The experimental data consist of $R_{AA}(p_T)$ values with uncertainties over the measured jet-momentum range. The proton–proton spectrum $\Phi_{pp}(p_T)$ is taken from PYTHIA~8 simulations at the same center-of-mass energy and is interpolated in logarithmic space, with a power-law extrapolation at high $p_T$.

\begin{figure}[htbp]
    \centering
    \includegraphics[width=0.4\textwidth]{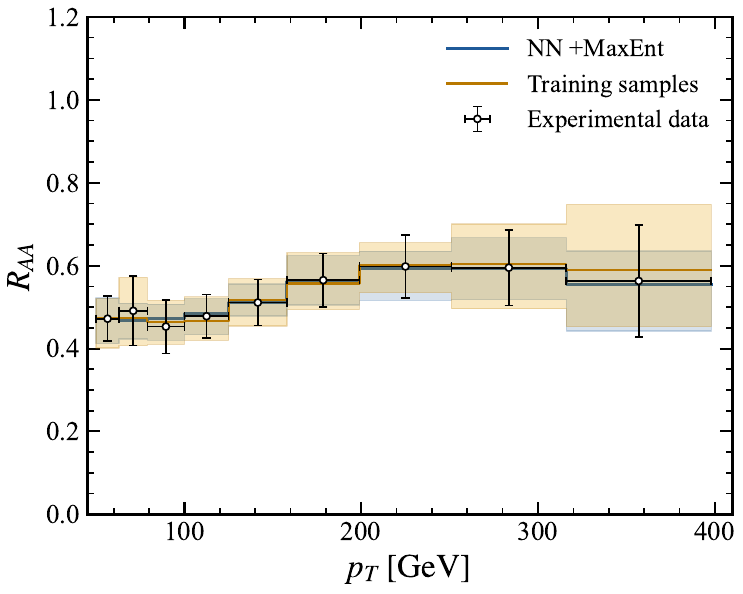}
    \caption{The nuclear modification factor \(R_{AA}\) in Pb+Pb collisions at \(\sqrt{s_{NN}}=2.76\)~TeV. Black circles: experimental data. Blue curve and band: NN+MaxEnt ensemble median and its central \(16\)--\(84\%\) interval. Orange curve and band: bootstrap training samples generated from the experimental \(R_{AA}\).}
    \label{fig:raa}
\end{figure}

Figure~\ref{fig:raa} compares the measured \(R_{AA}\) with the values obtained by passing the reconstructed energy-loss distributions through the forward map. The black circles with error bars denote the experimental data entering the training, while the blue curve and band represent the NN+MaxEnt ensemble median and central 16--84\% interval. Their agreement is therefore a consistency check of the inverse reconstruction, rather than an independent prediction of \(R_{AA}\).

\begin{figure}[htbp]
    \centering
    \includegraphics[width=\columnwidth]{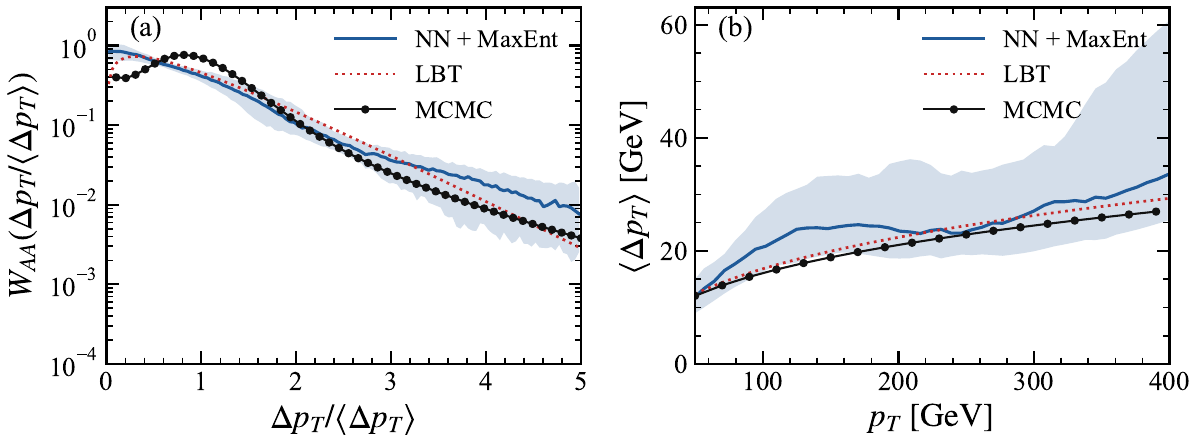}
    \caption{(a) Scaled energy-loss distribution as a function of $u=\Delta p_{\mathrm{T}}/\langle\Delta p_{\mathrm{T}}\rangle$. The NN+MaxEnt result (blue line and band) is obtained from a uniformly weighted marginalization over the selected initial-momentum interval; the band denotes the central $16$--$84\%$ ensemble interval. Black points and line: MCMC result from Ref.~\cite{He2018}. Red dotted line: LBT result from Ref.~\cite{He2018}. The comparison is intended to be qualitative. (b) Conditional mean energy loss $\langle\Delta p_T\rangle(x)$ as a function of the initial jet momentum $x$.}
    \label{fig:eloss}
\end{figure}

Figure~\ref{fig:eloss}(a) shows the scaled energy-loss distribution as a
function of \(u=\Delta p_{\mathrm{T}}/\langle\Delta p_{\mathrm{T}}\rangle\),
together with the Bayesian MCMC extraction and LBT calculation of
Ref.~\cite{He2018}. The NN+MaxEnt result is obtained by Monte Carlo
marginalization of the conditional distribution \(q_\theta(z|x)\):
\(x\) is sampled uniformly over the chosen initial-momentum range,
\(z\sim q_\theta(z|x)\) is then drawn, and the resulting
\(\Delta p_{\mathrm{T}}=xz\) samples are rescaled by their mean. The
resulting curve therefore represents a uniformly weighted marginal over
the selected initial-momentum interval, rather than the conditional
distribution at a fixed \(x\).

The reconstructed distribution exhibits a broad, long-tailed structure
without assuming a Gamma functional form. Its qualitative trend is
compatible with the MCMC and LBT results, while differences are visible
at small \(u\). These differences should not be interpreted as evidence
for a unique microscopic low-energy-loss structure. As is clear from
Eq.~(\ref{eq:raa_integral}), the mapping from the inclusive
one-dimensional \(R_{AA}(p_T)\) to the two-variable conditional density
\(q_\theta(z|x)\) is underdetermined. The detailed reconstructed shape
therefore reflects the combined influence of the information contained
in the measured \(R_{AA}\), the forward operator, the entropy and
smoothness regularization, and the marginalization prescription. For
this reason, the comparison with the MCMC and LBT curves is intended to
be qualitative.

Figure~\ref{fig:eloss}(b) shows the conditional mean energy loss,
\[
\langle\Delta p_{\mathrm{T}}\rangle(x)
=
\int d\Delta p_{\mathrm{T}}\,
\Delta p_{\mathrm{T}}
W_{AA}(\Delta p_{\mathrm{T}}|x),
\]
as a function of the initial jet momentum \(x\). The NN ensemble gives
a median \(\langle\Delta p_{\mathrm{T}}\rangle\simeq11.8\)~GeV at
\(x=50\)~GeV, with a central \(16\)--\(84\%\) interval of approximately
\(9.0\)--\(15.0\)~GeV, and remains broadly consistent with the MCMC and
LBT trends over the calculated range.

\section{Summary and outlook}
\label{sec:sum}
We have developed a neural-network, maximum-entropy-regularized framework
for reconstructing probability distributions and conditional densities
from limited observables connected to them through differentiable forward
models. The method uses a positive neural representation without imposing
a fixed analytic family and combines observable constraints with entropy
regularization and, when needed, explicit smoothness penalties. Because
the constraints are implemented through finite loss terms, NN+MaxEnt is
best viewed as an entropy-regularized neural implementation of the
maximum-entropy principle rather than an exact solution of the classical
constrained MaxEnt problem.

Gaussian, Poisson, and mixed-Poisson closure tests demonstrate that the
framework can recover the dominant features of distributions compatible
with the imposed low-order constraints. A physics-motivated test based on
rescaled fRG cumulants further shows that NN+MaxEnt accurately reproduces
the imposed cumulants and yields effective multiplicity distributions
close to those obtained with the conventional Lagrange-multiplier MaxEnt
construction. Since the volume factor, experimental acceptance, and
baryon-to-proton mapping are not fixed, this part of the analysis is a
closure benchmark rather than a quantitative prediction of a measured
proton-number distribution.

We then applied the same framework to the inverse reconstruction of a
conditional jet-energy-loss distribution from single-inclusive jet
\(R_{AA}\) data in central (0--10\%) Pb+Pb collisions at
\(\sqrt{s_{NN}}=2.76\)~TeV. The reconstructed energy-loss distributions
reproduce the measured suppression through the forward map without
assuming a Gamma or other fixed functional form. At an initial jet
momentum \(x=50\)~GeV, the conditional mean energy loss is
\(\langle\Delta p_T\rangle\simeq11.8\)~GeV, with a central
\(16\)--\(84\%\) ensemble interval of approximately
\(9.0\)--\(15.0\)~GeV. The scaled energy-loss profile is qualitatively
consistent with the Bayesian MCMC extraction and LBT calculation of
Ref.~\cite{He2018}.

The jet inversion is intrinsically underdetermined: an inclusive
one-dimensional \(R_{AA}(p_T)\) does not uniquely fix the conditional
density \(W_{AA}(\Delta p_T|x)\). Accordingly, the detailed reconstructed
shape reflects the measured data together with the forward operator and
the adopted entropy, smoothness, and marginalization prescriptions. The
reported ensemble band propagates pointwise experimental uncertainties
and initialization variability, but does not represent the full
systematic uncertainty associated with these modeling choices. Within
these limitations, NN+MaxEnt provides a flexible and less ansatz-dependent
strategy for regularized distribution reconstruction in heavy-ion
physics. Future work can incorporate experimental covariance information
and more complete systematic uncertainties, and extend the same approach
to higher-dimensional inverse problems.

\section*{Acknowledgments}

F.P. L acknowledges support from the NSFC under grant No.~12325507, No.~12547102, and No.~12147101, and the National Key Research and Development Program of China under grant No. 2022YFA1604900. This work was supported by the National Natural Science Foundation of China (NSFC) under Grant Nos.\ 12075098, 12435009, 12225503 and in part by the China Postdoctoral Science Foundation under Grant No.\ 2025M783370. Financial support was also provided by the self-determined research funds of Central China Normal University (CCNU) from the colleges' basic research and operation of the Ministry of Education (MOE). We gratefully acknowledge the extensive computing resources provided by the Nuclear Computing Center of Central China Normal University.
For further information about the data and code used in this study, please contact the authors via email.

Declaration of generative AI and AI-assisted technologies in the writing process

During the preparation of this work the author(s) used deepseek in order to polish the language. After using this tool/service, the author(s) reviewed and edited the content as needed and take(s) full responsibility for the content of the publication.

\appendix
\section{Sensitivity to regularization parameters}
\label{sec:lambda}

Table~\ref{tab:lambda_scan} summarizes a one-at-a-time sensitivity test
of the regularization parameters using the extracted mean energy loss as
a diagnostic. Within this reduced test, the mean energy loss is nearly
unchanged when the entropy or smoothness terms are removed and remains
stable under moderate variations of the normalization weight. In
contrast, increasing \(\lambda_{\rm ent}\) from \(10^{-3}\) to
\(10^{-2}\) produces a substantial shift, indicating over-regularization.
The production choice \(\lambda_{\rm ent}=10^{-3}\) therefore corresponds
to a weak entropy bias for this observable. This scan is not intended to
establish the full regularization uncertainty of the reconstructed
functional shape. The bootstrap uncertainty bands shown in the main text
do not include this regularization dependence.

\begin{table}[htbp]
\centering
\caption{One-at-a-time sensitivity scan of the regularization
coefficients. The scan uses a reduced setup with
\(N_{\rm boot}=3\), 800 epochs, and \(N_{\rm quad}=80\), and is intended
as a limited robustness check of the mean energy loss around the production choice
\((\lambda_N,\lambda_{\rm ent},\lambda_z,\lambda_x)
=(20,10^{-3},5\times10^{-4},10^{-4})\).}
\label{tab:lambda_scan}
\begin{tabular}{lccccc}
\hline
case & $\lambda_N$ & $\lambda_{\rm ent}$ & $\lambda_z$ & $\lambda_x$
& $\langle\Delta p_T\rangle_{x=200\,{\rm GeV}}$ [GeV] \\
\hline
baseline & 20 & $10^{-3}$ & $5\times10^{-4}$ & $10^{-4}$ & 24.76 \\
$\lambda_N=10$ & 10 & $10^{-3}$ & $5\times10^{-4}$ & $10^{-4}$ & 24.38 \\
$\lambda_N=50$ & 50 & $10^{-3}$ & $5\times10^{-4}$ & $10^{-4}$ & 26.74 \\
$\lambda_{\rm ent}=0$ & 20 & 0 & $5\times10^{-4}$ & $10^{-4}$ & 24.57 \\
$\lambda_{\rm ent}=10^{-2}$ & 20 & $10^{-2}$ & $5\times10^{-4}$ & $10^{-4}$ & 37.95 \\
$\lambda_z=0$ & 20 & $10^{-3}$ & 0 & $10^{-4}$ & 24.77 \\
$\lambda_x=0$ & 20 & $10^{-3}$ & $5\times10^{-4}$ & 0 & 24.78 \\
\hline
\end{tabular}
\end{table}

\bibliography{ref}

\end{document}